# Anharmonic thermodynamic properties and phase boundary across the post-perovskite transition in MgSiO$_3$

Zhen Zhang[1] and Renata M. Wentzcovitch[1,2,3,*]

[1]*Department of Applied Physics and Applied Mathematics, Columbia University, New York, NY 10027, USA.*

[2]*Department of Earth and Environmental Sciences, Columbia University, New York, NY 10027, USA.*

[3]*Lamont–Doherty Earth Observatory, Columbia University, Palisades, NY 10964, USA.*

[*]Corresponding author: rmw2150@columbia.edu

**Abstract**

   To address the effects of lattice anharmonicity across the perovskite to post-perovskite transition in MgSiO$_3$, we conduct calculations using the phonon quasiparticle (PHQ) approach. The PHQ is based on *ab initio* molecular dynamics and, in principle, captures full anharmonicity. Free energies in the thermodynamic limit ($N \to \infty$) are computed using temperature-dependent quasiparticle dispersions within the phonon gas model. Systematic results on anharmonic thermodynamic properties and phase boundary are reported. Both the local density approximation (LDA) and the generalized gradient approximation (GGA) calculations are performed to provide confident constraints on these properties. Anharmonic effects are demonstrated by comparing results with those obtained using the quasiharmonic approximation (QHA). The inadequacy of the QHA is indicated by its overestimation of thermal expansivity and thermodynamic Grüneisen parameter and its converged isochoric heat capacity in the high-temperature limit. The PHQ phase boundary has a Clapeyron slope ($dP/dT$) that increases with temperature. This result contrasts with the nearly zero curvature of the QHA phase boundary. Anharmonicity bends the phase boundary to lower temperatures at high pressures. Implications for the double-crossing of the phase boundary by the mantle geotherm are discussed.

## I. INTRODUCTION

MgSiO$_3$ perovskite (Pv) with the *Pbnm* space group, also known as bridgmanite, is the most abundant mineral in the Earth's lower mantle (LM), composing about 75 vol% of this region [1]. It undergoes a structural phase transition to MgSiO$_3$ post-perovskite (PPv) with the *Cmcm* space group under the lowermost mantle conditions, i.e., above ~125 GPa and ~2500 K [2-4]. The lowermost mantle with several hundred kilometers of depth, i.e., the D" layer, is one of the least understood regions in the Earth's interior [5]. It is the region where the D" seismic discontinuity is observed [6,7], and the PPv phase transition may be a reasonable interpretation of such seismic discontinuity. Previous experimental measurements for the phase boundary in pure MgSiO$_3$ were conducted up to ~3000 K, yielding significant discrepancies in the Clapeyron slopes, $dP/dT$ (~4.7–11.5 MPa/K), and the transition pressures (~113–131 GPa at 2500 K) [2,3,5,8,9]. Numerous theoretical predictions for the phase boundary in MgSiO$_3$ were also carried out, giving various Clapeyron slopes (~7.5–9.9 MPa/K) and also a significant discrepancy in the transition pressures (~108–128 GPa at 2500 K) [3,4,9,10]. Previous theoretical studies of this phase boundary [3,4,9,10] used the quasiharmonic approximation (QHA), which disregards intrinsic lattice anharmonic effects. Therefore, the role of anharmonic effects on the phase boundary should still be investigated.

To capture and understand the role of anharmonic effects on the PPv phase transition, we use the phonon quasiparticle (PHQ) approach [11,12] to compute the anharmonic phonon dispersions and vibrational free energies. The PHQ method relies on the atomic trajectories obtained from *ab initio* molecular dynamics (AIMD) simulations to compute the mode-projected velocity autocorrelation function (VAF). This method, in principle, treats anharmonicity exactly to all orders in perturbation theory. It deals with lattice anharmonicity by extracting phonon quasiparticle properties, i.e., renormalized phonon frequencies and phonon lifetimes, from the mode-projected VAFs. Unlike the QHA, in which phonon frequencies depend on volume only, PHQ accounts for full anharmonicity and produces phonon frequencies explicitly temperature-dependent. Then the vibrational free energy can be computed using these renormalized temperature-dependent frequencies. The direct free energy method, such as thermodynamic integration (TI) [13], can also address lattice anharmonicity. However, to approach the thermodynamic limit ($N \to \infty$), performing TI using AIMD with a sufficiently large supercell is computationally unaffordable. In the present approach, the renormalized frequencies are Fourier interpolated on a sufficiently dense **q** point mesh to overcome finite-size effects on the vibrational free energy. Then the vibrational entropy and free energy are computed in the thermodynamic limit from the interpolated renormalized frequencies, i.e., the anharmonic phonon dispersions, within the phonon gas model (PGM) [14,15].

The PHQ approach was proposed and verified in Pv [11], of which irregular temperature-induced frequency shifts [16-18] observed in the Raman spectrum were successfully reproduced. In this study, we compute the anharmonic phonon dispersions of Pv and PPv using both the local density approximation (LDA) [19] and the Perdew-Burke-Ernzerhof generalized gradient approximation (PBE-GGA) [20]. Isochoric AIMD simulations at a series of $V, T$ conditions are conducted to cover LM conditions, i.e., $23 < P < 135$ GPa, $2000 < T < 4000$ K. Systematic results on anharmonic thermodynamic properties and phase boundary across the PPv phase transition in $MgSiO_3$ are reported. Anharmonic effects on the thermodynamic properties and phase boundary are demonstrated by comparing to results obtained with the QHA.

## II. METHOD

We define a phonon quasiparticle numerically by the mode-projected VAF [11,12],

$$\langle V_{\mathbf{q}s}(0) \cdot V_{\mathbf{q}s}(t)\rangle = \lim_{\tau \to \infty} \frac{1}{\tau} \int_0^\tau V_{\mathbf{q}s}^*(t') V_{\mathbf{q}s}(t' + t) dt', \quad (1)$$

where

$$V_{\mathbf{q}s}(t) = \sum_{i=1}^N \sqrt{M_i} \mathbf{v}_i(t) e^{i\mathbf{q}\cdot\mathbf{R}_i} \cdot \hat{\mathbf{e}}_{\mathbf{q}s} \quad (2)$$

is the mass-weighted mode-projected velocity for normal mode (**q**, s). **q** is the phonon wave vector, and $s$ indexes the $3n$ phonon branches of an $n$-atom primitive cell. $M_i$, $\mathbf{R}_i$, and $\mathbf{v}_i$ ($i = 1, ..., N$) are the atomic mass, the atomic equilibrium coordinate, and the atomic velocity computed by AIMD simulations of an $N$-atom supercell, respectively. $\hat{\mathbf{e}}_{\mathbf{q}s}$ is the harmonic phonon polarization vector determined by the harmonic phonon calculations, and **q** is commensurate with the supercell size. For a well-defined phonon quasiparticle, the VAF can be phenomenologically described as an exponentially decaying cosine function,

$$\langle V_{\mathbf{q}s}(0) \cdot V_{\mathbf{q}s}(t)\rangle = A_{\mathbf{q}s} \cos(\widetilde{\omega}_{\mathbf{q}s} t) e^{-\Gamma_{\mathbf{q}s} t}, \quad (3)$$

where $A_{\mathbf{q}s}$ is the oscillation amplitude, $\widetilde{\omega}_{\mathbf{q}s}$ is the renormalized phonon frequency, and $\Gamma_{\mathbf{q}s}$ is the phonon linewidth inversely proportional to the lifetime, $\tau_{\mathbf{q}s} = 1/(2\Gamma_{\mathbf{q}s})$ [11,21]. The corresponding power spectrum,

$$G_{\mathbf{q}s}(\omega) = \left|\int_0^\infty \langle V_{\mathbf{q}s}(0) \cdot V_{\mathbf{q}s}(t)\rangle e^{i\omega t} dt\right|^2, \quad (4)$$

should have a Lorentzian line shape with a single peak at $\widetilde{\omega}_{\mathbf{q}s}$ and a linewidth of $\Gamma_{\mathbf{q}s}$ [11,21]. In this study, the obtained $\widetilde{\omega}_{\mathbf{q}s}$ are used to compute the anharmonic thermodynamic properties and phase boundary, while $\tau_{\mathbf{q}s}$ can be used to evaluate the lattice thermal conductivity [22]. To overcome the finite-size effects on the vibrational free energy and obtain the free energy in the thermodynamic limit, the anharmonic phonon spectrum is further calculated via Fourier interpolation for $\widetilde{\omega}_{\mathbf{q}s}$ [11,12]. Within the PGM, the vibrational entropy formula with temperature-dependent anharmonic phonon spectrum is applicable [11,12,21],

$$S_{\text{vib}}(T) = k_B \sum_{\mathbf{q}s}[(n_{\mathbf{q}s} + 1)\ln(n_{\mathbf{q}s} + 1) - n_{\mathbf{q}s}\ln n_{\mathbf{q}s}], \quad (5)$$

where $n_{\mathbf{q}s} = [\exp(\hbar\widetilde{\omega}_{\mathbf{q}s}(T)/k_B T) - 1]^{-1}$. $\widetilde{\omega}_{\mathbf{q}s}(T)$ at arbitrary temperatures were obtained by fitting $\widetilde{\omega}_{\mathbf{q}s}$ at several temperatures and constant volume to a second-order polynomial in $T$ [11,23,24]. Then for insulators such as Pv and PPv, the Helmholtz free energy $F(T)$ at any $T$ can be obtained by integrating the vibrational entropy [11,25,26],

$$F(T) = E_0 + \frac{1}{2}\sum_{\mathbf{q}s} \hbar\omega_{\mathbf{q}s} - \int_0^T S_{\text{vib}}(T')dT'. \quad (6)$$

$E_0$ is the static energy and $\frac{1}{2}\sum_{\mathbf{q}s} \hbar\omega_{\mathbf{q}s}$ is the zero-point energy, where $\omega_{\mathbf{q}s}$ is the harmonic phonon frequency, i.e., the phonon frequency at zero temperature. The latter two terms on the right-hand side compose the vibrational free energy.

We performed AIMD simulations using the projected-augmented wave method [27] as implemented in VASP [28]. The electron exchange-correlation functional (XC) was treated by both the LDA [19] and the PBE [20]. Pv and PPv were simulated with 160-atom supercells ($2 \times 2 \times 2$) and 180-atom supercells ($3 \times 3 \times 2$), respectively, both with a $\Gamma$ **k** point sampling. The supercells are sufficiently large to converge the phonon quasiparticle properties as reported by previous studies [11,22,29]. For each phase and each XC, MD simulations were carried out in the *NVT* ensemble at a series of 5 volumes and a series of 6 temperatures between 300 and 5000 K controlled by the Nosé thermostat [30,31]. Each MD ran for over 50 ps with a time step of 1 fs. To approach the thermodynamic limit, anharmonic phonon spectra were evaluated with much denser **q** meshes ($16 \times 16 \times 8$ for Pv and $20 \times 20 \times 10$ for PPv), sufficiently dense to converge the vibrational free energy. Static calculations were conducted using the well-converged $4 \times 4 \times 4$ and $8 \times 8 \times 4$ **k** point sampling for Pv and PPv, respectively.

### III. RESULTS AND DISCUSSION

Temperature-dependent anharmonic phonon dispersions at constant volume are displayed in Figs. 1(a)–1(d). Temperature-induced frequency shifts are discernible but small for Pv and PPv, meaning that both phases are weakly anharmonic. Using such anharmonic phonon frequencies interpolated on a dense **q**-mesh, the Helmholtz free energy $F(T)$ at constant volume was computed by integrating the vibrational entropy via Eqs. (5) and (6). Then, at each temperature, the isothermal equation of state (EOS) was obtained by fitting $F(V)$ at several volumes to a third-order finite strain expansion [23,24]. The resulting $F(V,T)$ are shown in Figs. 1(e) and 1(f).

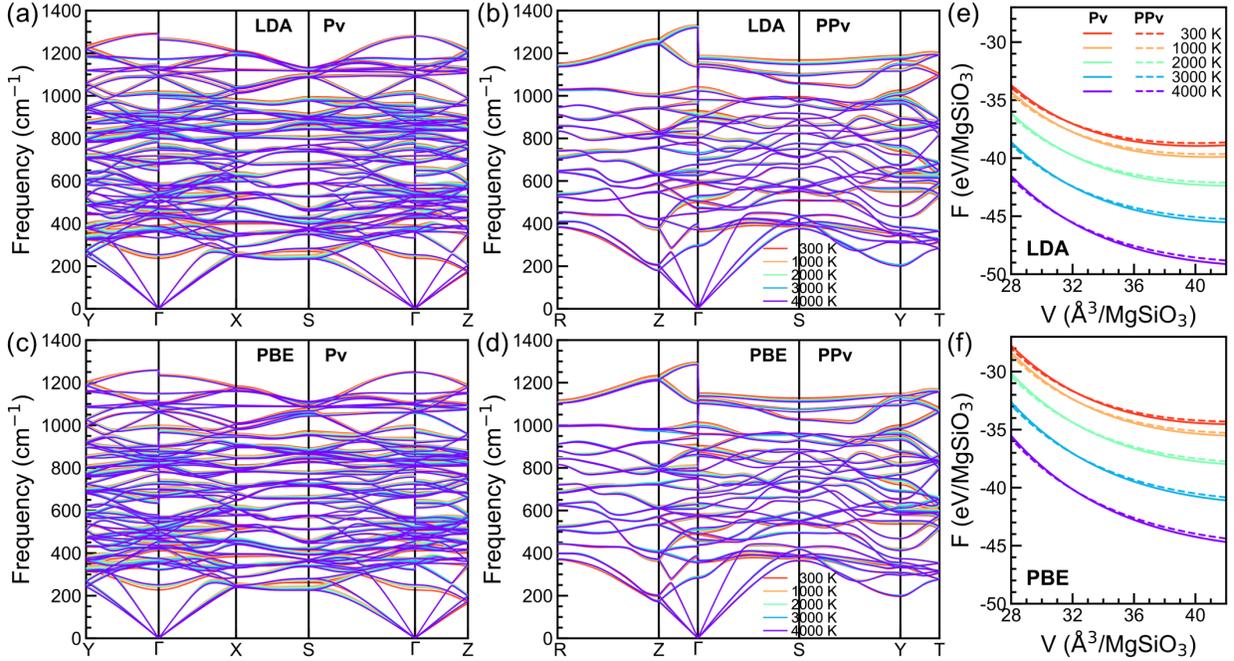

FIG. 1. (a)–(d) Temperature-dependent anharmonic phonon dispersions obtained by the (a)(b) LDA and (c)(d) PBE for (a)(c) Pv and (b)(d) PPv. Results are shown at their respective volumes with a static pressure of 120 GPa. (e)(f) Helmholtz free energy $F(V,T)$ vs. volume at different temperatures. Results are obtained by the (e) LDA and (f) PBE for Pv (solid curves) and PPv (dashed curves).

The obtained temperature-dependent third-order Burch-Murnaghan EOS parameters, i.e., equilibrium volume ($V_0$), bulk modulus ($K_0$), and pressure derivative of the bulk modulus ($K_0'$) are displayed in Figs. 2(a)–2(f). For both phases and both XCs, as temperature increases, $K_0$ decrease whereas $K_0'$ increase. For both phases, $V_0$ obtained by the LDA are smaller than those obtained by the PBE, while $K_0$ obtained by the LDA are larger than those obtained by the PBE. For both XCs, $K_0$ of Pv are larger than those of PPv, while $K_0'$ of Pv are smaller than those of PPv. The calculated EOS parameters at 300 K are summarized and compared to reported experimental measurements in TABLE I. The equilibrium volumes at room temperature predicted by the LDA are smaller than experiments for both Pv and PPv, while results by the PBE are significantly larger than experiments. The bulk moduli at room temperature predicted by the LDA agree well with experiments for both phases, whereas predictions by the PBE are smaller than experiments. Therefore, the LDA reproduces the EOS better than the PBE for both phases. This is further justified by comparing the volumes at higher pressures and temperatures, as shown in Figs. 2(g) and 2(h). The volumes at 300 and 2000 K predicted by the LDA are smaller than experimental measurements. In contrast, the volumes predicted by the PBE are significantly larger than in experiments.

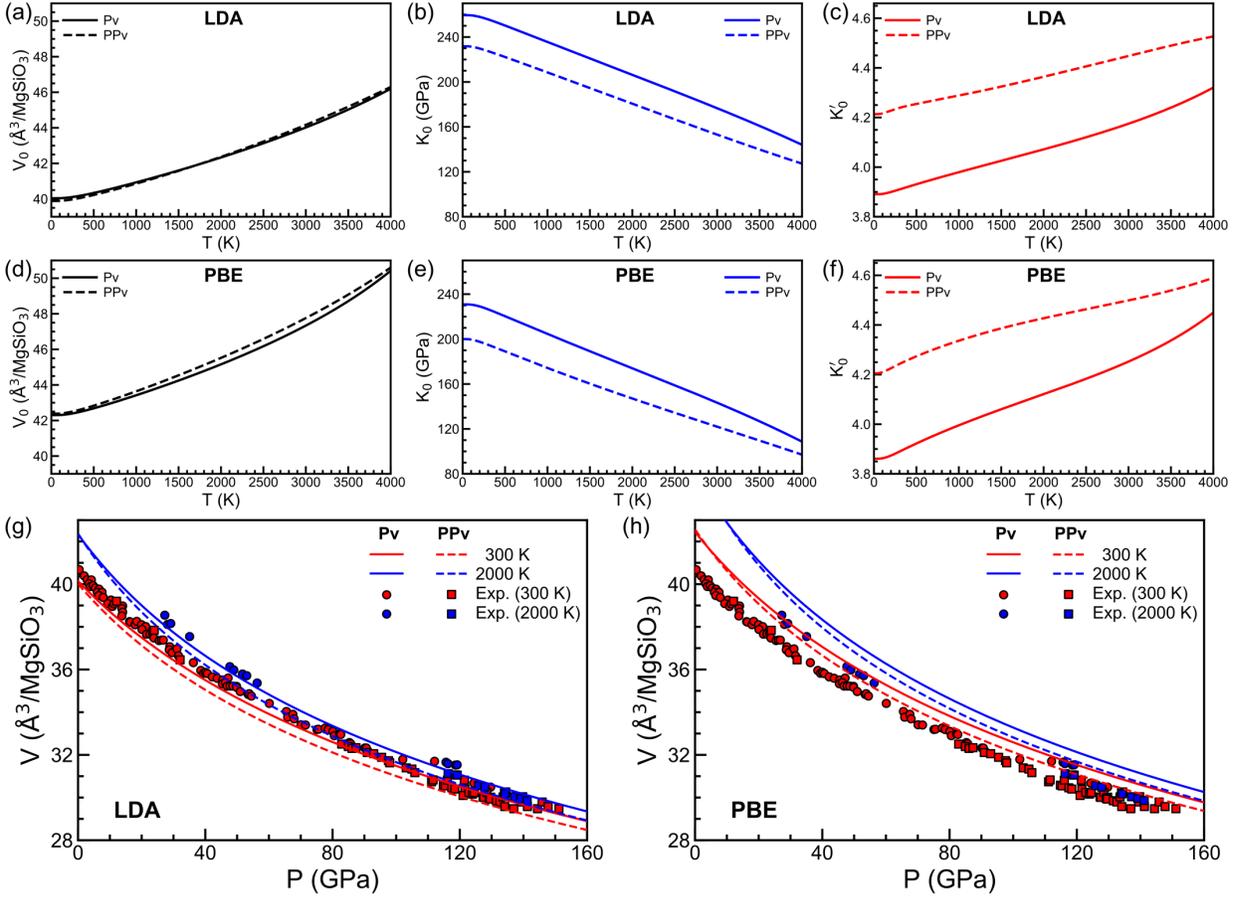

FIG. 2. (a)–(f) Temperature-dependent third-order Burch-Murnaghan EOS parameters, i.e., equilibrium volume ($V_0$), bulk modulus ($K_0$), and pressure derivative of the bulk modulus ($K_0'$). Results are obtained with the (a)–(c) LDA and (d)–(f) PBE for Pv (solid curves) and PPv (dashed

curves). (g)(h) EOS at 300 (red) and 2000 K (blue) compared with reported experimental measurements (circles for Pv and squares for PPv). Results are obtained with the (g) LDA and (h) PBE for Pv (solid curves) and PPv (dashed curves). Pv experimental data are taken from [32-41] at 300 K and [37,38,40] at 2000 K. PPv experimental data are taken from [2,35,40,42,43] at 300 K and [40,43] at 2000 K.

TABLE I. Calculated EOS parameters at 300 K compared to experimental measurements for Pv [33-35,44-49] and PPv [35,42,43].

|     |      | $V_0$ (Å³/MgSiO₃) | $K_0$ (GPa) | $K_0'$ |
|-----|------|-------------------|-------------|--------|
| Pv  | LDA  | 40.15             | 255.72      | 3.91   |
|     | PBE  | 42.44             | 226.47      | 3.89   |
|     | Exp. | 40.58–40.83       | 246–272     | 3.65–4.00 |
| PPv | LDA  | 40.02             | 227.43      | 4.24   |
|     | PBE  | 42.56             | 194.91      | 4.24   |
|     | Exp. | 40.55–41.23       | 219–248     | 4.00 (fixed) |

With fully anharmonic $F(V,T)$, anharmonic thermodynamic quantities are readily calculated. Pressure, thermal expansivity ($\alpha$), isothermal bulk modulus ($K_T$), isochoric heat capacity ($C_V$), thermodynamic Grüneisen parameter ($\gamma$), adiabatic bulk modulus ($K_S$), and isobaric heat capacity ($C_P$) are obtained from $P = -\left(\frac{\partial F}{\partial V}\right)_T$, $\alpha = \frac{1}{V}\left(\frac{\partial V}{\partial T}\right)_P$, $K_T = -V\left(\frac{\partial P}{\partial V}\right)_T$, $C_V = T\left(\frac{\partial S}{\partial T}\right)_V$, $\gamma = \frac{V\alpha K_T}{C_V}$, $K_S = K_T(1 + \gamma\alpha T)$, and $C_P = C_V(1 + \gamma\alpha T)$, respectively. Systematic results on the anharmonic thermodynamic properties for both phases using both XCs are shown in Fig. 3. For both XCs, $\gamma$ and $C_P$ of Pv are generally larger than those of PPv, especially at high temperatures and low pressures. For both phases, the LDA yields smaller $\alpha$ but larger $K_T$ and $K_S$ than the PBE. All reported quantities are essential in geodynamic modeling. In particular, $\alpha$ and $\gamma$ are helpful indicators of the relative importance of anharmonicity compared to the QHA results [50-53]. $K_T$ and $K_S$, along with other thermoelastic properties, are essential in interpreting seismic tomography [54]. $C_V$ and $C_P$ are needed in theory and experiments to evaluate the lattice thermal conductivity [55,56].

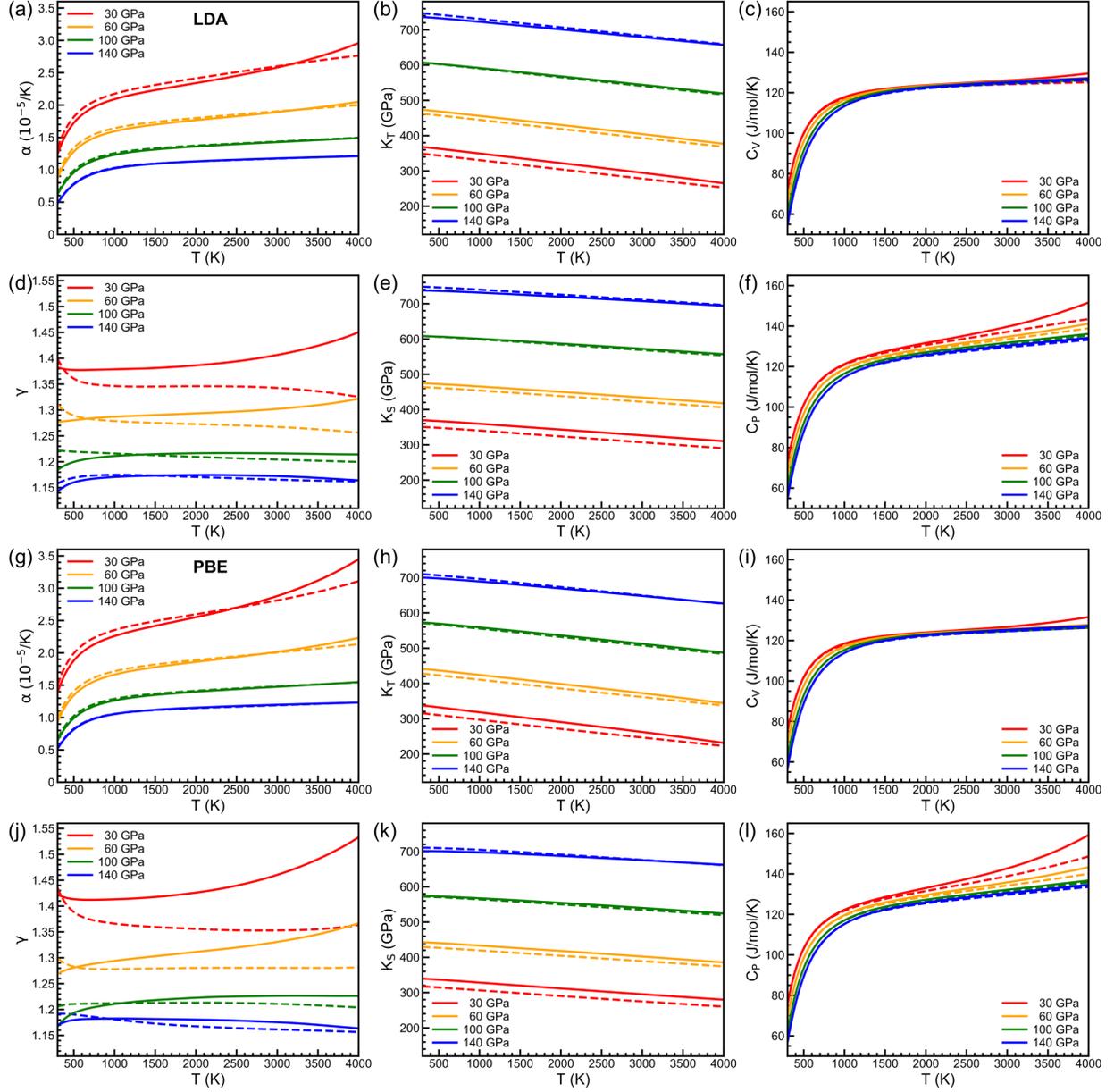

FIG. 3. Thermodynamic quantities, i.e., thermal expansivity ($\alpha$), isothermal bulk modulus ($K_T$), isochoric heat capacity ($C_V$), thermodynamic Grüneisen parameter ($\gamma$), adiabatic bulk modulus ($K_S$), and isobaric heat capacity ($C_P$) vs. temperature at different pressures. Results are obtained by the (a)–(f) LDA and (g)–(l) PBE for Pv (solid curves) and PPv (dashed curves).

Anharmonic effects on the thermodynamic properties are further analyzed by comparing $\alpha$, $\gamma$, and $C_V$ at high temperatures, e.g., 4000 K with results obtained by the QHA, as displayed in Fig. 4. It is known that the overestimation of $\alpha$ compared to experiments, especially at high temperatures and low pressures, is the fingerprint of the QHA's shortcomings [50-53]. For Pv, $\alpha$ predicted by the QHA agrees very well with that by the PHQ, even down to 30 GPa at 4000 K. This means the QHA works quite well for Pv. For PPv, in contrast, the QHA overestimates $\alpha$ compared to the PHQ results, and the overestimation becomes more and more significant at low pressures. Note that in the QHA, the temperature effects are accounted for by extrinsic volumetric effects only, i.e., quasiharmonic thermal expansivity [57]. Whereas in the PHQ, anharmonicity is treated as the intrinsic temperature dependence of phonon frequencies and is, in principle, fully included by the method. Our direct comparison between the QHA and PHQ results justifies the statement about $\alpha$ as an indicator of the QHA's shortcomings [50-53].

$\gamma$ is another important indicator of the importance of anharmonicity [51]. The inadequacy of the QHA is also seen in its overestimation of $\gamma$ compared to experiments for other minerals [51,53]. For Pv, the QHA overestimates $\gamma$ compared to the PHQ, yet the difference remains roughly constant as pressure decreases. For PPv, however, the overestimation of $\gamma$ by the QHA becomes significant at low pressures. Such observation is in line with that for $\alpha$, both of which lead to the conclusion that the QHA works better for Pv than PPv and the QHA becomes inadequate at high temperatures and low pressures.

The insufficiency of the QHA and the anharmonic effects captured by the PHQ is further seen in the comparison of $C_V$. At high temperatures, $C_V$ obtained using the QHA converges to an upper limit of $3nk_\mathrm{B}$. This is because $3nk_\mathrm{B}$ is the high-temperature limit within the Debye model [58] for harmonic crystals with temperature-independent frequencies. In contrast, $C_V$ obtained using the temperature-dependent frequencies from the PHQ can exceed this limit at high temperatures [23,59,60]. Note that $C_V$ obtained by the QHA at high temperatures is subject to the mathematical limit. The relatively larger difference between the QHA and PHQ results for Pv's $C_V$ does not contradict the previous conclusion that the QHA works better for Pv than PPv.

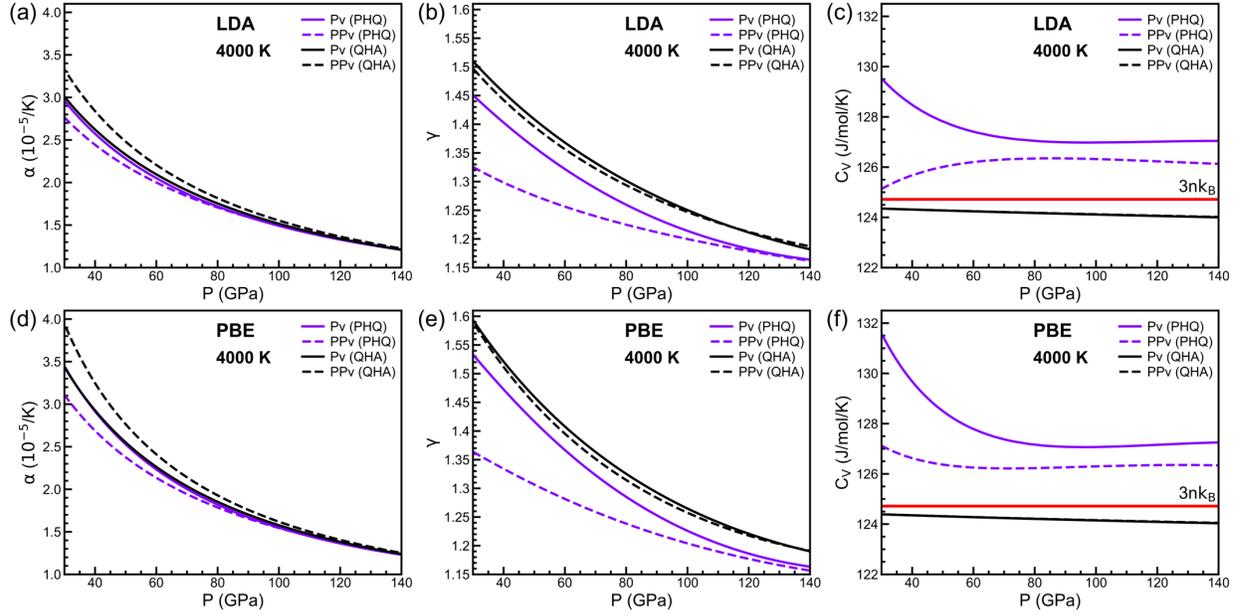

FIG. 4. Comparison of (a)(d) $\alpha$, (b)(e) $\gamma$, and (c)(f) $C_V$ computed with the PHQ approach (purple) and the QHA (black) vs. pressure at 4000 K. Results are obtained by the (a)–(c) LDA and (d)–(f) PBE for Pv (solid curves) and PPv (dashed curves). The red horizontal lines label the high-temperature limit, $3nk_B$, within the Debye model.

The Helmholtz free energy is then converted into the Gibbs free energy, $G = F + PV$. By comparing $G$ of Pv and PPv, phase boundaries using both XCs are obtained and shown in Fig. 5. QHA results are also displayed to demonstrate anharmonic effects on the phase boundary. Previously reported experimental measurements and theoretical predictions are also shown. The discrepancies among the reported experimental results are significant, with transition pressures ranging from ~113–131 GPa at 2500 K and Clapeyron slopes ranging from ~4.7–11.5 MPa/K [2,3,5,8,9]. This is mainly caused by the different pressure scales used in the measurements [5,8], e.g., Au scales [61-64], MgO scale [65], and Pt scale [66]. The discrepancies among the reported theoretical results are also significant, with transition pressures ranging from ~108–128 GPa at 2500 K and Clapeyron slopes ranging from ~7.5–9.9 MPa/K [3,4,9,10]. Although the LDA gives better EOS than the GGA, the GGA (PBE-GGA [3,9,57] or WC-GGA [10]) has been reported to reproduce better polymorphic phase boundaries than the LDA [3,9,10,57].

We obtain transition pressures to be 110 and 117 GPa at 2500 K by the LDA and the PBE-GGA, respectively. LDA and PBE-GGA phase boundaries in this study agree relatively well and fall in between the two boundaries predicted by the same functionals by Oganov and Ono [3,9]. Our transition pressures are considerably smaller than those reported by Tsuchiya *et al.* [4] for the same functionals. This may be caused by either the use of an under-converged $4 \times 4 \times 2$ **k** point sampling for PPv or the use of different software, i.e., Quantum ESPRESSO [67] by Tsuchiya *et al.* [4]. Our calculated PBE-GGA phase boundary agrees better with experiments [2,3,5,8,9] than the LDA, which is in line with previous reports [3,9,10,57]. All previous calculations [3,4,9,10] obtained thermodynamic properties using the QHA, and the predicted phase boundaries above ~500 K have zero curvature. The PHQ phase boundary shows a curvature and deviates more and more from the QHA boundary with increasing temperature. The calculated Clapeyron slopes by both methods and both XCs are summarized in TABLE II. The QHA gives a temperature-independent 8.0 MPa/K by averaging the LDA and PBE results, in agreement with previous calculations [3,4,10]. The PHQ Clapeyron slope increases with temperature. The average values of the slopes by the two XCs are 7.6 MPa/K at 1000 K, 9.0 MPa/K at 2500 K, and 10.3 MPa/K at 4000 K. The large values of the Clapeyron slope at high temperatures are in better agreement with the measurements by Hirose *et al.* [5,8] using the MgO scale [65].

The last column of TABLE II summarizes the calculated transition temperatures at the core-mantle boundary (CMB) by both methods and both XCs. The QHA LDA phase boundary beyond 5000 K was linearly extrapolated. The solid-liquid phase of $MgSiO_3$ [68-71] was not considered and is not in the scope of this study. By averaging the LDA and PBE results, the PHQ gives a transition temperature of 4660 K at the CMB. The accumulated errors in the QHA lead to an average overestimation of 510 K, i.e., 5170 K at the CMB. Our PHQ result is in fairly good agreement with a recently reported experimental transition temperature of 4800 K at the CMB [72]. Such temperatures are much higher than the present-day CMB temperature of ~4000 K (3600–4300 K [73-75]). Therefore, the geotherm [76] crosses the calculated phase boundary only once at ~2500 K. The double-crossing hypothesis [77] states that the back transformation from PPv to Pv may occur within the D" layer above the CMB. Our results show the double-crossing hypothesis does not hold for pure $MgSiO_3$. Nevertheless, anharmonicity bends the phase boundary to lower temperatures at high pressures, and anharmonic effects facilitate the double-crossing scenario [77]. Additional shifts [78,79] and broadening [78-84] of the phase boundary in mantle occurring alloys among $MgSiO_3$, $Al_2O_3$, $FeSiO_3$, and $Fe_2O_3$ still need to be carefully and consistently addressed to shed more light on the double-crossing issue.

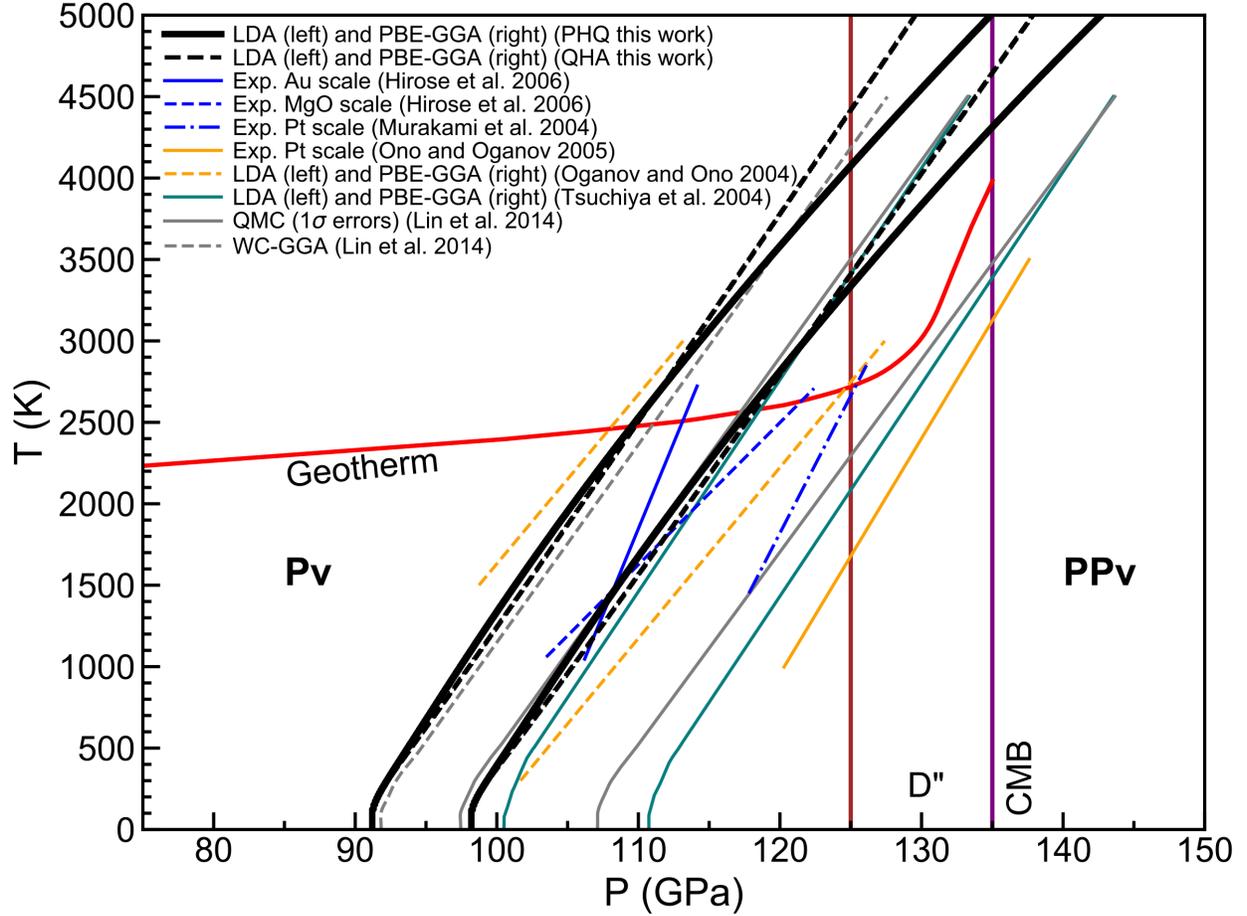

FIG. 5. Pv-PPv phase boundaries obtained in this work with the PHQ (solid black curves) and the QHA (dashed black curves) under the LM conditions are compared with reported experimental measurements [2,3,5,8,9] and previous theoretical predictions [3,4,9,10] in pure MgSiO$_3$. The two solid gray curves by the quantum Monte Carlo (QMC) show the boundaries of a 1$\sigma$ error band [10]. The red curve shows the geotherm [76]. The vertical brown and purple lines show the onset of the D" region and the CMB, respectively.

TABLE II. Calculated Clapeyron slopes at different temperatures and transition temperatures at the CMB.

|  |  | Clapeyron slope (MPa/K) | | | Transition temperature (K) |
| --- | --- | --- | --- | --- | --- |
|  |  | at 1000 K | at 2500 K | at 4000 K | at the CMB |
| PHQ | LDA | 7.6 | 9.0 | 10.3 | 5010 |
|  | PBE | 7.7 | 9.0 | 10.4 | 4310 |
| QHA | LDA | 7.9 | 7.9 | 7.9 | 5690 |
| | PBE | 8.1 | 8.1 | 8.1 | 4650 |

## IV. CONCLUSIONS

Using the phonon quasiparticle (PHQ) approach [11,12], we computed temperature-dependent anharmonic phonon quasiparticle dispersions of MgSiO$_3$ perovskite (Pv) and post-perovskite (PPv). Both the LDA [19] and PBE [20] were used to provide confident constraints on the thermodynamic properties and phase boundary for the Pv-PPv transition. Fully anharmonic free energies were calculated in the thermodynamic limit ($N \to \infty$) within the phonon gas model [14,15]. Thermal equation of state, thermal expansivity ($\alpha$), thermodynamic Grüneisen parameter ($\gamma$), isothermal ($K_T$) and adiabatic ($K_S$) bulk moduli, isochoric ($C_V$) and isobaric ($C_P$) heat capacities, and phase boundary were then obtained. Comparing PHQ and QHA results, we see that $\alpha$ and $\gamma$ are overestimated by the QHA, especially for PPv at high temperatures and low pressures. At high temperatures, QHA $C_V$ converges to the high-temperature limit of $3nk_B$, while the PHQ $C_V$ can exceed this limit. The PPv phase transition pressure is 114 GPa at 2500 K by averaging the LDA and PBE results. Anharmonic effects produce a phase boundary with non-zero curvature above 500 K. The Clapeyron slope ($dP/dT$) increases with temperature, e.g., 7.6 MPa/K at 1000 K, 9.0 MPa/K at 2500 K, and 10.3 MPa/K at 4000 K. The transition temperature at the core-mantle boundary (CMB) is 4660 K, which is 510 K lower than the QHA prediction, yet still much higher than the expected present-day CMB temperature. Hence, the geotherm [76] crosses the phase boundary only once at ~2500 K, and the double-crossing phenomenon [77] should not happen in pure MgSiO$_3$. Additional Al and Fe alloying effects on the Pv-PPv transition must still be considered before the geotherm's double-crossing of the phase boundary in natural mantle composition can be addressed [78-84].

## ACKNOWLEDGMENTS

This work was primarily funded by the US Department of Energy Grant DE-SC0019759 and in part by the National Science Foundation (NSF) award EAR-1918126. This work used the Extreme Science and Engineering Discovery Environment (XSEDE), USA, supported by the NSF Grant ACI-1548562. Computations were performed on Stampede2, the flagship supercomputer at the Texas Advanced Computing Center (TACC), the University of Texas at Austin, generously funded by the NSF through Grant ACI-1134872.